# Kondo effect in under-doped n-type superconductors

Tsuyoshi Sekitani, Michio Naito*, Noboru Miura

*Institute for Solid States Physics, University of Tokyo, Kashiwanoha, Kashiwa-shi, Chiba, 277-8581, Japan*

*\*NTT Basic Research Laboratories, Morinosato-Wakamiya, Atsugi-shi, Kanagawa, 243-0198, Japan*

## Abstract

We present high-field magnetotransport properties of high-quality single-crystalline thin films of heavily under-doped nonsuperconducting $(La,Ce)_2CuO_4$, $(Pr,Ce)_2CuO_4$, and $(Nd,Ce)_2CuO_4$. All three materials show identical behavior. They are metallic at high temperatures and show an insulating "upturn" at low temperatures. The insulating upturn has a $\log T$ dependence, but saturates toward the lowest temperatures. Notably, the insulating upturn tends to be suppressed by applying magnetic fields. This negative magnetoresistance has a $\log B$ dependence, and its anisotropy shows non simple behavior. We discuss these findings from the viewpoints of Kondo scattering and also two-dimensional weak localization, and demonstrate Kondo scattering as a more plausible explanation. The Kondo scatters are identified as $Cu^{2+}$ spins in the $CuO_2$ planes.

74.25.Fy, 74.72.Jt, 74.76.Bz, 75.70.-i

It has been claimed that the normal state of high-$T_c$ cuprate superconductors is "anomalous". Especially there has been considerable controversy as to whether the normal ground state is metallic or insulating. To clarify this issue, the transport properties of the low-temperature normal state have been investigated on various cuprates by suppressing superconductivity with high magnetic fields. Boebinger, Ando et al. performed extensive studies on hole-doped $La_{2-x}Sr_xCuO_4$ (LSCO) and $Bi_2Sr_{2-x}La_xCuO_{6+y}$, and observed unusual insulating behavior, which they claimed to be a generic normal-state property of under-doped cuprates [1–4]. Recently, Fournier et al. investigated the normal state of electron-doped $(Pr,Ce)_2CuO_4$ (PCCO) in detail, and pointed out many similarities to hole-doped LSCO in spite of the "apparent" electron-hole doping asymmetry as frequently mentioned [5, 6]. These results seem to indicate that cuprates, regardless of being the hole- or electron-doped, commonly show an "insulator-metal" crossover as a function of the doping level. In the under-doped regime, the resistivity shows an insulating "upturn" ($d\rho/dT<0$) at low temperatures, which has a $\log T$ dependence in many cases. There has been much speculation with respect to its origin, but a clear explanation has not yet been reached. In order to unveil the nature of this $\log T$ insulating upturn, we have performed systematic magnetotransport experiments on electron-doped $(Nd,Ce)_2CuO_4$ (NCCO), PCCO, and $(La,Ce)_2CuO_4$ (LCCO) using pulsed magnetic fields up to 50 T [7, 8]. In this article we present the results for heavily under-doped nonsuperconducting specimens. The purpose of using non-superconducting samples is to avoid the influence of superconductivity and thereby to observe the normal-state transport down to zero magnetic field. Our results strongly indicate that the origin of the low-temperature "insulating" behavior is the Kondo scattering by magnetic moments of $Cu^{2+}$.

Magnetotransport experiments were performed on high-quality $c$-axis oriented NCCO, PCCO, and LCCO films at temperatures from 1.5 K to 300 K. The films were grown by MBE on either $SrTiO_3$ (001) or $KTaO_3$ (001) substrates. The thickness of the films was $\sim 1000$ Å. All of the films, for which we show the results in this article, are heavily under-doped and nonsuperconducting. In order to show the quality of our films, the $T_c$ and resistivity values for our typical optimum doped films as well as the heavily under-doped films used for this study are summarized in Table I. The requirements to prepare such high-quality of films are stringent cation stoichiometry adjustment and careful removal of apical oxygen without phase decomposition. The details of our film growth were described in ref [9–11].

High magnetic fields up to 50 T were produced by a pulse magnet, which was energized by a capacitor bank of 900 kJ (5 kV or 10 kV). The resistivity was measured by the standard four-probe method with electrodes formed by Ag or Au evaporation. A DC current of the order of 1∼10 $\mu$A was supplied in plane.

The in-plane resistivities in different magnetic fields applied parallel to the $c$-axis are shown in Fig. 1. The insets show the linear-scale replots of the zero-field data. The behaviors for all NCCO, PCCO, and LCCO films are similar to one another. The resistivity shows metallic behavior ($d\rho/dT > 0$) with a $T^2$ dependence down to the resistivity-minimum temperature ($T_{min}$). Below $T_{min}$, the resistivity increases as the temperature is lowered (we call this behavior as an "upturn" in this article). This insulating upturn has a $\log T$ dependence, but saturates toward the lowest temperatures. Furthermore the upturn tends to be suppressed by applying magnetic fields. Similar results have recently been reported for nonsuperconducting PCCO films by Fournier $et\ al.$ [6].

Figure 2 shows the in-plane resistivity as a function of magnetic field ($B$) at various temperatures. Here the field is applied parallel to the $c$-axis. The data are plotted against $\log B$ in order to see the field dependence of the negative magnetoresistance. It was confirmed that there is no heating effect due to eddy currents in our experiments since the data recorded from both up- and down-sweeps of the pulsed magnetic field nearly coincide. The behaviors for all NCCO, PCCO, and LCCO are again almost identical. Negative magnetoresistance appears at temperatures below the resistivity minimum ($T_{min}$) and becomes more prominent with decreasing temperature. It has a $\log B$ dependence, but saturates toward $B = 0$ T. [footnote: In hole-doped cuprates such as LSCO, negative magnetoresistance in in-plane transport have been reported by other groups [1, 12]. Especially, Preyer $et\ al.$ observed isotropic negative magnetoresistance in high-quality single crystal of LSCO [12].]

In order to demonstrate the anisotropy of the negative magnetoresistance, Fig. 3 shows the magnetoresistance curves against $\log B$ for different field directions. The angle dependences of the magnetoresisitance are depicted in Fig. 4. The angle ($\theta$) in this figure is defined with respect to the $c$-axis. The curves, except for $\theta = 90°$, almost fall into the same line, indicating weak anisotropy. For $\theta = 90°$, however, the curve deviates from this trend line. The dissimilar behavior for $\theta = 90°$, where $B$ is parallel to the layer, is not well understood, however, at least in some part may be related to the complicated magnetic structure in the charge reservoir blocks ($Ln_2O_2$ layers) since different behavior is observed for different Ln.

The two main features, the log-$T$ dependent insulating upturn and the log-$B$ dependent negative magnetoresistance, which have been observed in the present experiment, can be explained either by localization or by Kondo scattering. The former possibility has been pointed out by several authors [5,6,13,14]. Calculations based on 2-dimensional (2D) weak localization predict, for simplest cases, a $\log T$ dependent divergence of resistivity toward $T = 0$ K and a $\log B$ dependent negative magnetoresistance. However, this possibility may be ruled out because the anisotropy of the negative magnetoresistance does not follow a cosine dependence. Moreover, in 2D weak localization, the coefficient ($\alpha$) for $\log T$ dependent conductivity and also for $\log B$ dependent magnetoconductivity per sheet should be material-independent universal values. However, we found that the observed $\alpha$ value depends on the doping level and also on the material. Figure 5 shows such an example, which shows the doping dependence of the coefficient for NCCO [7,8]. The coefficient, $\alpha$, changes significantly from 0.9 to 14.6 as $x$ varies from 0.086 (heavily underdope) to 0.146 (optimum).

The weak anisotropy of the observed negative magnetoresistance suggests that the observed phenomena should be spin related in origin. Hence, next we discuss our experimental results on the basis of the Kondo scattering. The Kondo effect, which arises from the exchange interaction between itinerant conduction electrons and localized spin impurities, leads to anomalous temperature dependences in various physical parameters due to the Fermi surface effect. The anomalous behaviors, in essence, originate from singlet formation between a conduction electron and a localized spin below the Kondo crossover temperature ($T_K$). With regard to the resistivity, a third order Born approximation for the spin inversion scattering gives rise to a $\log T$ dependent upturn. This $\log T$ dependent resistivity does not diverge toward 0 K, but has a finite maximum value (unitarity-limit scattering) at $T$ = 0 K, which differs from the behavior predicted by 2D weak localization. High magnetic fields act to suppress the spin inversion (or, equivalently, to dissociate Kondo singlets) and thereby give negative magnetoresistance, which is isotropic in ideal cases. The negative magnetoresistance has a $\log B$ dependence in the intermediate field region, and saturates both toward $B = 0$ T and toward $B > B_K$ ($B_K$: Kondo crossover field). This simplest Kondo description can be applied to Al and noble metals (Cu, Ag, and Au) containing $3d$ magnetic impurities like Mn and Fe, and also to some rare earth compounds containing $4f$ magnetic impurities. One typical material for the latter category is (La,Ce)B$_6$, for which extensive magnetotransport data are available [15]. The magnetotransport properties observed here for (Ln,Ce)$_2$CuO$_4$ qualitatively agree with the above Kondo predictions, and furthermore look very similar to those for (La,Ce)B$_6$. In simplest Kondo systems, negative magnetoresistance should be isotropic. In actual cases, however, weak but finite anisotropy has been reported for anisotropic materials [16–19]. The anisotropic negative magnetoresistance has been understood as an anisotropic $g$-value due to the crystalline electrical field. A similar explanation may also be applied to our cases.

We give some quantitative discussions below. First we compare the experimental temperature and magnetic-field dependences with theoretical predictions for the Kondo resistivity. To extract the Kondo resistivity $\rho_K(T, B)$, we assume that the resistivity can be decomposed into three parts,

$$\rho(T, B) = \rho_0 + \rho_i(T) + \rho_K(T, B) \tag{1}$$

where $\rho_0$ represents temperature-independent impurity scattering and $\rho_i$ a high-temperature $T^2$ component (empirical, the origin of this $T^2$ dependnce is not known). Here we neglect conventional orbital (positive) magnetoresistance that might give a slight $B$ dependence to $\rho_0$ and to $\rho_i$. With a standard fitting procedure for subtracting $\rho_0$ and $\rho_i$ as shown in Fig. 6(a), we can get $\rho_K(T, B=0\ T)$ in Fig. 6(b) and $\rho_K(T=1.5\ K, B)$ in Fig. 6(c). The approximate theoretical expression (so called the "KMHZ" or generalized "Hamann" formula [20]) for the Kondo resistivity $\rho_K(T, B)$ is given as

$$\rho_K(T,B) = \frac{\rho_u}{2}\left\{1 - \frac{f(T,B)}{\sqrt{f^2(T,B) + \pi^2\left[S(S+1) + \frac{1}{4}tanh^2\frac{\beta g\mu_B B}{2}\right]}}\right\} \quad (2)$$

$$f(T,B) = \ln\frac{T}{T_K} + \mathrm{Re}\psi\left[\frac{1}{2} + \frac{\beta g\mu_B B}{2\pi i}\right] - \psi\left[\frac{1}{2}\right]$$

where $\rho_u$ represents the unitarity-limit resistivity, and $\psi(z)$ denotes the digamma function ($\psi(z)$ has a logarithmic dependence in the asymptotic limit of $z \to \infty$) [21,22]. This formula is based on the Suhl-Nagaoka theory for the Kondo effect [23–25]. In Fig. 6(b), the zero-field data are compared with the prediction (dashed line) by eq. (2) with $T_K = 15.7$ K, 18.7 K, and 17.9 K for LCCO, PCCO, and NCCO, respectively. The agreement is fair for $T/T_K \geq 1$. However, the experimental data deviates from eq. (2) for $T/T_K < 1$. This discrepancy may be due to inapplicability of eq. (2) to $T/T_K \ll 1$. For $T/T_K \ll 1$, most of the existing data for $\rho_K(T, B=0\ T)$ for typical Kondo materials can be better described by the "empirical" formula [26,27], which is given as

$$\rho_K(T) = \frac{\rho_u}{2}\left\{1 - \frac{ln\,[(T^2+\theta^2)/T_K^2]^{\frac{1}{2}}}{\pi\,[S(S+1)]^{\frac{1}{2}}}\right\} \quad (3)$$

where $\ln(\theta/T_K) = -\pi\,[S(S+1)]^{\frac{1}{2}}$, namely $\theta = 0.066T_K$ for $S = 1/2$ [25]. Our experimental data for $T/T_K \leq 1$ is also in a good agreement with this "empirical" formula, which is indicated by a solid line in Fig. 6(b). In spite of this good agreement for the temperature dependence, the magnetic field dependence $\rho_K(T=1.5\ K, B)$ does not agree with eq. (2) as is shown in Fig. 6(c). This discrepancy might be ascribed to the approximate of the KMHZ formula [21,22]. The Kondo crossover temperature $T_K$ and the Kondo crossover field $B_K$ as determined $k_B T_K = S g_{eff} \mu_B B_K$ with $S = 1/2$ and $g_{eff} = 2$ are summarized in Table I.

Finally we discuss the origin of the Kondo scatterer. $Nd^{3+}$ has a paramagnetic spin moment, and $Pr^{3+}$ can also have a paramagnetic spin moment at high magnetic fields although the ground state is nonmagnetic. Therefore, the interaction between conduction electrons and spin moments of $Nd^{3+}$ or $Pr^{3+}$ might be a candidate for the Kondo scattering. Actually, Maiser *et al.* pointed out the possible interaction between conduction electrons and spin moments of $Nd^{3+}$ ions at very low temperatures [28]. However, essentially similar behavior is also observed in LCCO, where $La^{3+}$ should not have a spin moment. So we can exclude the possibility of $Nd^{3+}$ or $Pr^{3+}$ spins as the Kondo scatterer. Therefore we explore the other possibility, i.e. $Cu^{2+}$ local spins in the $CuO_2$ plane. Regarding the magnetism in the $CuO_2$ planes for electron-doped $(Ln,Ce)_2CuO_4$, there have been many studies in the past, although the results have not been well converging. One early study by $\mu$SR for NCCO indicated that antiferromagnetism persists up to $x \sim 0.14$, above which superconductivity suddenly sets in [29]. However, it is now well established that interstitial apical oxygen tends to stabilize antiferromagnetic correlation, and may modify the intrinsic magnetism in the T' compounds. Therefore different "recipes" to remove apical oxygen may give different sample properties. [footnote: This can be noticed by the facts that there exist very large sample dependence in the resistivity upturn and also the residual resistivity value reported in the past.] The intrinsic magnetic phase diagram for electron-doped $(Ln,Ce)_2CuO_4$ cannot be reached without complete removal of interstitial apical oxygen. According to the "old" magnetic phase diagram mentioned above [29], $Cu^{2+}$ should be antiferromagnetically ordered at low temperatures in all of the films studied here. If this *were* the case, single-impurity Kondo scattering approximation would not work. As an experimental fact, however, our experimental results can be well described by simple single-impurity Kondo theories. This fact may bring the "old" magnetic phase diagram into doubt. We suggest that a very small number of $Cu^{2+}$ Kondo impurities are induced by residual apical oxygen.

In summary, our magnetotransport experiments on heavily under-doped nonsupercon-ducting $(La,Ce)_2CuO_4$, $(Pr,Ce)_2CuO_4$, and $(Nd,Ce)_2CuO_4$ thin films have shown that the anomalous low-temperature transport common in these compounds is most likely governed by Kondo scattering by $Cu^{2+}$ spins in the $CuO_2$ planes. Our results may lead to a better understanding of the doping-induced "insulator-metal" crossover in cuprates.

We are obliged to Dr.Takashi Hayashi, Dr.Akio Tsukada, Dr.Kazuhito Uchida, Dr.Yasuhiro H.Matsuda, Satoru Ikeda, Douglas King, and Prof. Fritz Herlach for technical support and valuable discussions. This work was supported by a Grant-In-Aid for Scientific Research from the Ministry of Education, Science, Sports and Culture, Japan.

# REFERENCES

[1] Y. Ando *et al.*, Phys. Rev. Lett. **75**, 4662 (1995).

[2] G. S. Boebinger *et al.*, Phys. Rev. Lett. **77**, 5417 (1996).

[3] Y. Ando *et al.*, Phys. Rev. Lett. **77**, 2065 (1996).

[4] S. Ono *et al.*, Phys. Rev. Lett. **85**, 638 (2000).

[5] P. Fournier *et al.*, Phys. Rev. Lett. **81**, 4720 (1998).

[6] P. Fournier *et al.*, Phys. Rev. B **62**, R11993 (2000).

[7] T. Sekitani *et al.*, Physica B **294-295**, 358 (2001).

[8] T. Sekitani *et al.*, J. Phys. Chem. Solids **63**, 1089 (2002).

[9] M. Naito *et al.*, Physica C **293**, 36 (1997).

[10] H. Sato, and M. Naito, Physica C **274**, 221 (1997).

[11] M. Naito, and M. Hepp, Jpn. J. Appl. Phys. **39**, L485 (2000).

[12] N. W. Preyer *et al.*, Phys. Rev. B **44**, 407 (1991).

[13] S. J. Hagen *et al.*, Phys. Rev. B **45**, 515 (1992).

[14] Y. Hidaka *et al.*, J. Phys. Soc. Jpn. **60**, 1185 (1991).

[15] K. Samwer, and K. Winzer, Z. Physik B **25**, 269 (1976).

[16] Y. Onuki *et al.*, J. Phys. Soc. Jpn. **54**, 304 (1985).

[17] Y. Onuki *et al.*, J. Phys. Soc. Jpn. **54**, 3562 (1985).

[18] G. M. Roesler *et al.*, Phys. Rev. B **45**, 12893 (1992).

[19] K. Satoh *et al.*, J. Phys. Soc. Jpn. **61**, 3267 (1992).

[20] D. R. Hamann, Phys. Rev. **158**, 570 (1967).

[21] H. Keiter *et al.*, Solid State Commun. **16**, 1247 (1975).

[22] H. Keiter, Z. Physik B **23**, 37 (1976).

[23] Y. Nagaoka, Phys. Rev. **138A**, 1112 (1965).

[24] R. More, and H. Suhl, Phys. Rev. Lett. **20**, 500 (1968).

[25] M. D. Daybell, in Magnetism, edited by H. Suhl (Academic Press, New York, 1973), pp. 121-147.

[26] K. Winzer, Solid State Commun. **16**, 521 (1975).

[27] E. W. Fenton, Phys. Rev. B **7**, 3144 (1973).

[28] E. Maiser *et al.*, Phys. Rev. B **56**, 12961 (1997).

[29] G. M. Luke *et al.*, Phys. Rev. B **42**, 7981 (1990).

# I  TABLES

| Sample | $x$ | $T_c$(K) | $T_K$(K) | $B_K$(K) | $T_{min}$(K) | $\rho_0(\mu\Omega cm)$ | $\rho_u(\mu\Omega cm)$ | $\rho_{R.T.}(\mu\Omega cm)$ |
|---|---|---|---|---|---|---|---|---|
| La$_{2-x}$Ce$_x$CuO$_4$ | 0.045 | — | 15.7 | 23.4 | 103 | 198 | 720 | 750 |
|  | 0.090 | 30 | — | — | — | 25 | — | 250 |
| Pr$_{2-x}$Ce$_x$CuO$_4$ | 0.098 | — | 18.7 | 27.8 | 59 | 110 | 159 | 327 |
|  | 0.135 | 25 | — | — | — | 15 | — | 200 |
| Nd$_{2-x}$Ce$_x$CuO$_4$ | 0.086 | — | 17.9 | 26.6 | 73 | 300 | 570 | 763 |
|  | 0.145 | 24 | — | — | — | 15 | — | 150 |

TABLE I: Important parameters for Ln$_{2-x}$Ce$_x$CuO$_4$ (Ln=Nd, Pr, La) films with heavily under- and optimally-doping : Ce content ($x$), the superconducting temperature ($T_c$), the Kondo temperature ($T_K$), the Kondo magnetic field ($B_K$), the resistivity minimum temperature ($T_{min}$), the residual resistivity ($\rho_0$), the unitarity-limit resistivity ($\rho_u$), and the resistivity at 300 K ($\rho_{R.T.}$).

## II FIGURES

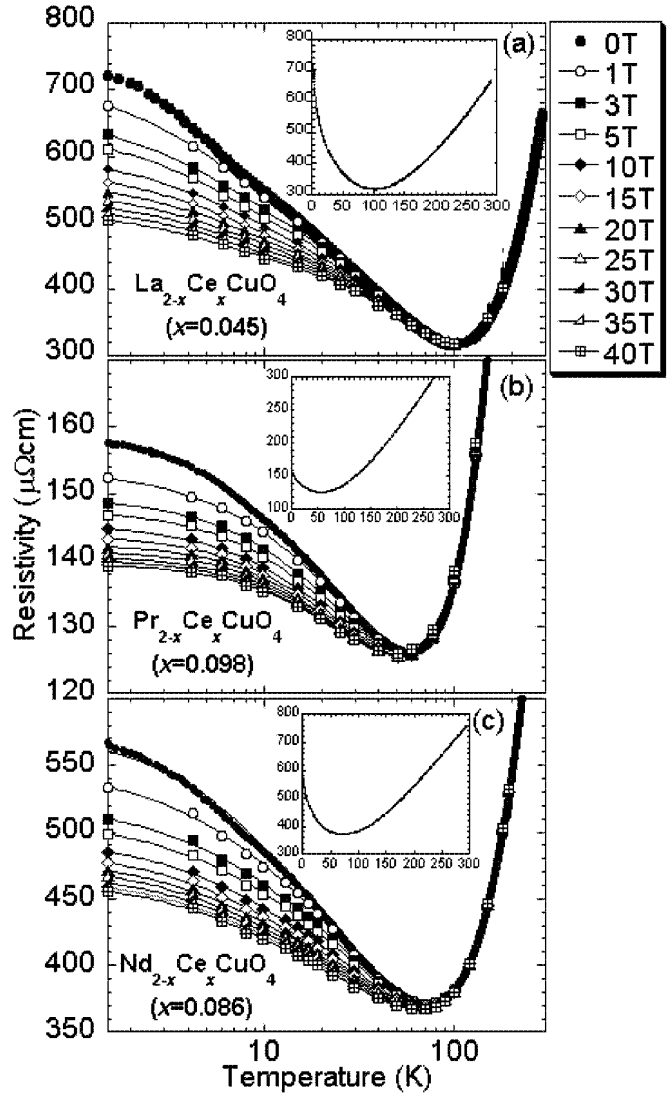

Fig. 1 / T. Sekitani et al.

FIG. 1: In-plane resistivities in magnetic fields as a function of $\log T$ for $(La,Ce)_2CuO_4$ (a), $(Pr,Ce)_2CuO_4$ (b), and $(Nd,Ce)_2CuO_4$ films (c). The insets show the linear-scale replots of the zero-field data.

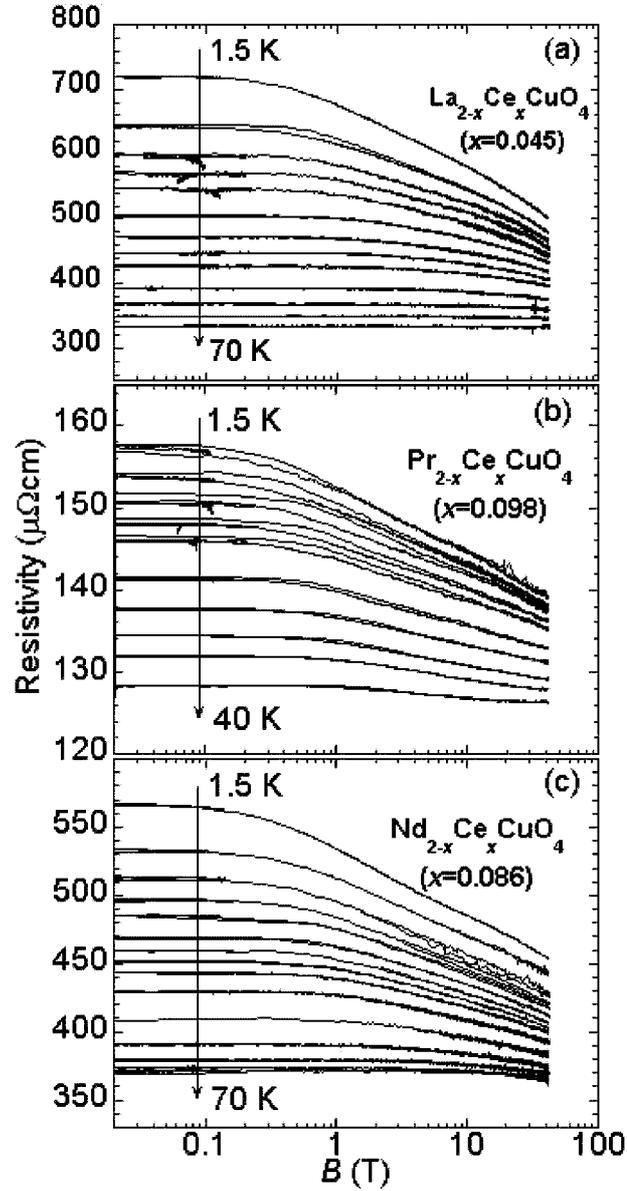

Fig. 2 / T. Sekitani et al.

FIG. 2: Magnetoresistance curves as a function of $\log B$ at different temperatures for (La,Ce)$_2$CuO$_4$ (a), (Pr,Ce)$_2$CuO$_4$ (b), and (Nd,Ce)$_2$CuO$_4$ films (c).

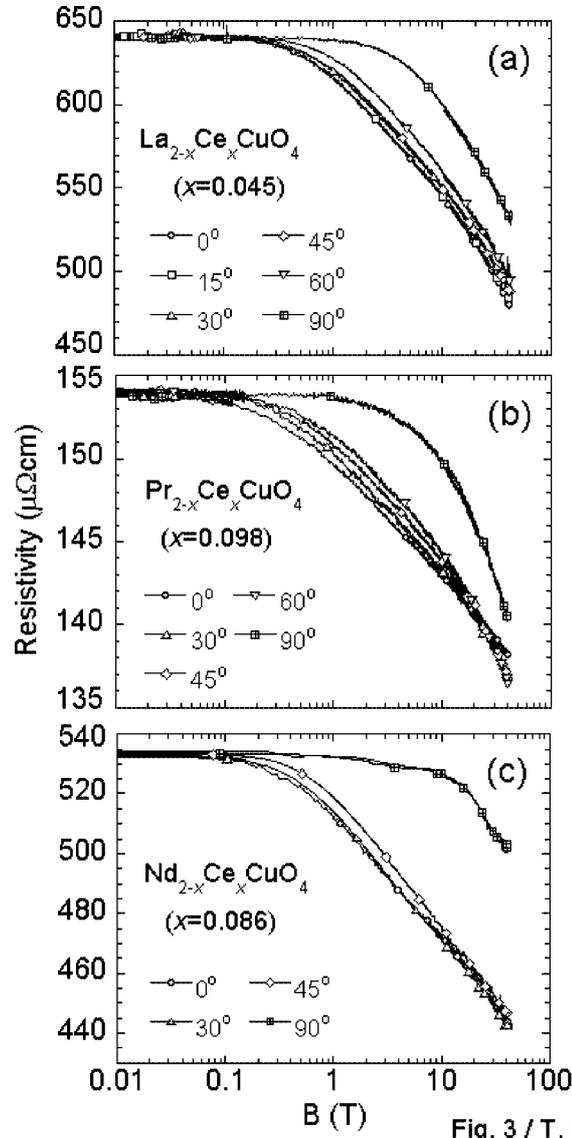

FIG. 3: Anisotropy of magnetoresistance at 4.2 K. The angle ($\theta$) of the magnetic field direction is defined with respect to the $c$-axis.

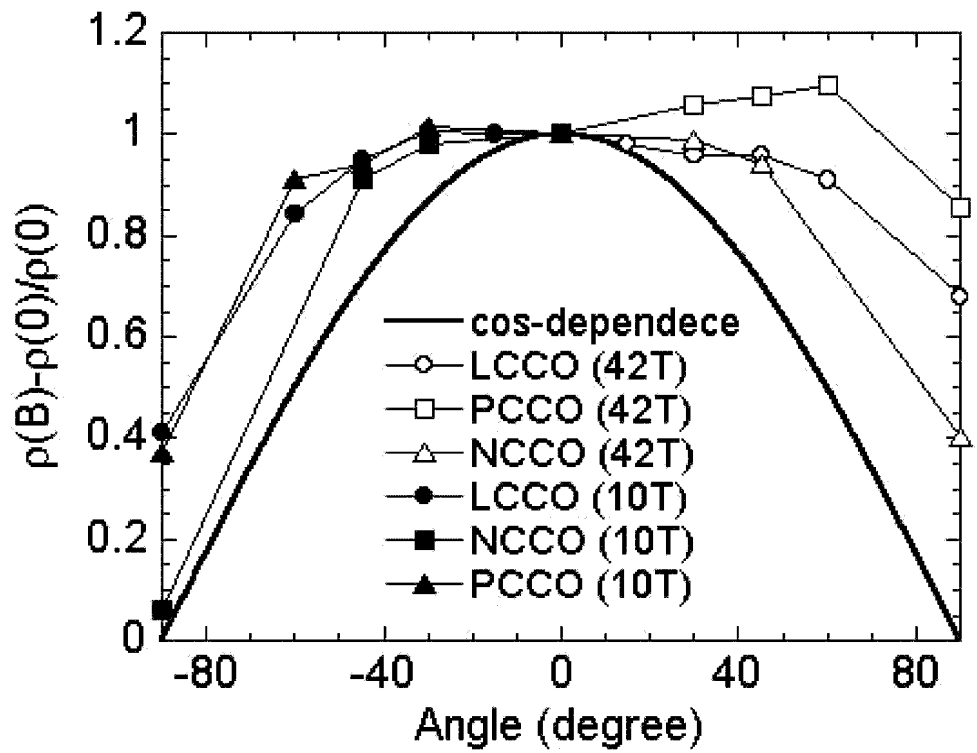

Fig. 4 / T. Sekitani *et al.*

FIG. 4: Comparison between the angular dependence of negative magnetoresistance at 4.2 K and a cosin dependence.

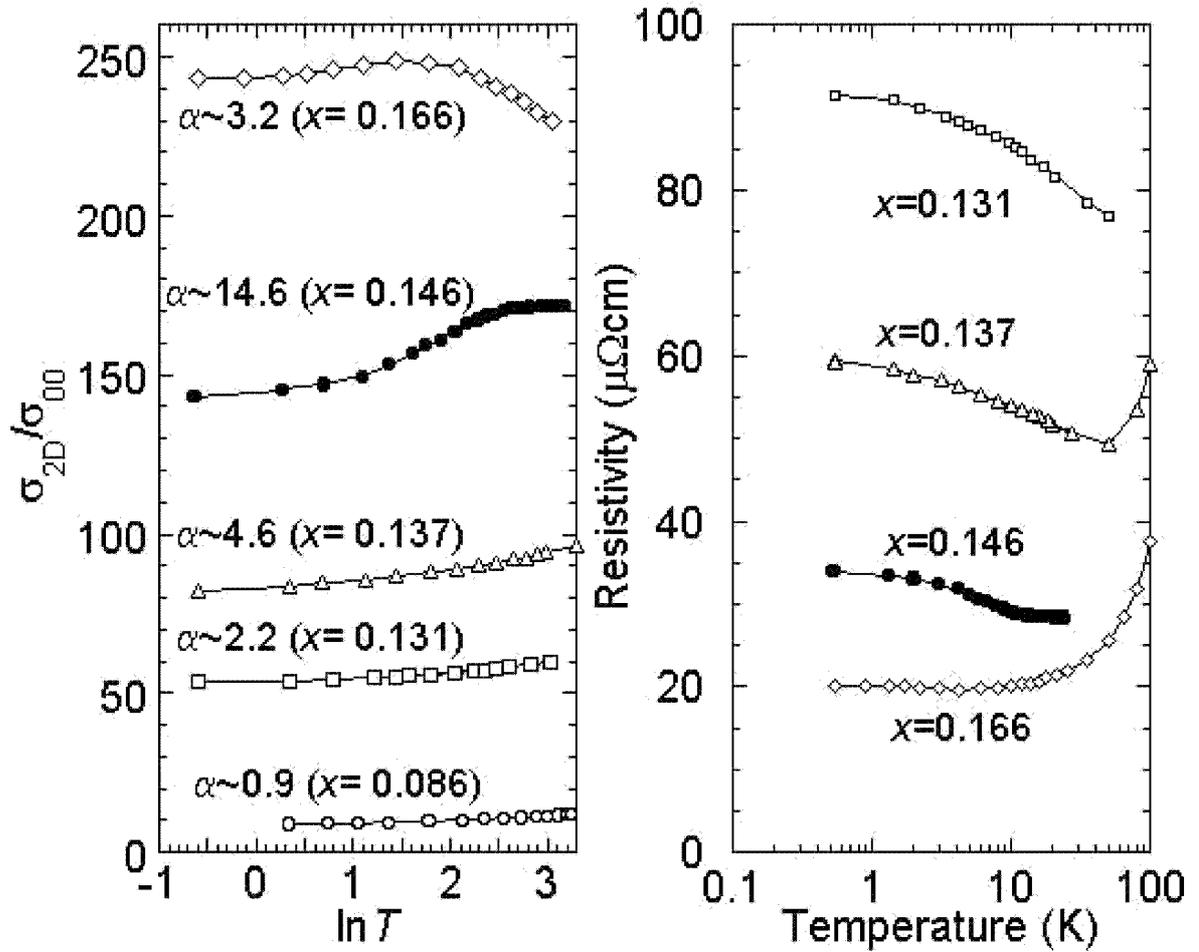

Fig. 5 / T. Sekitani et al.

FIG. 5: Doping dependence of upturn in NCCO films ($x$=0.086(under-doping) to 0.166(over-doping)) by suppressing the superconductivity with high magnetic fields at low temperatures [right], and the normalized values of the coefficient ($\alpha$) of $\log T$ ($\sigma_{2D} = d/\rho$, $\sigma_{00} = (e^2/\pi h)$, $\sigma_{2D}/\sigma_{00} = \alpha lnT$) [left].

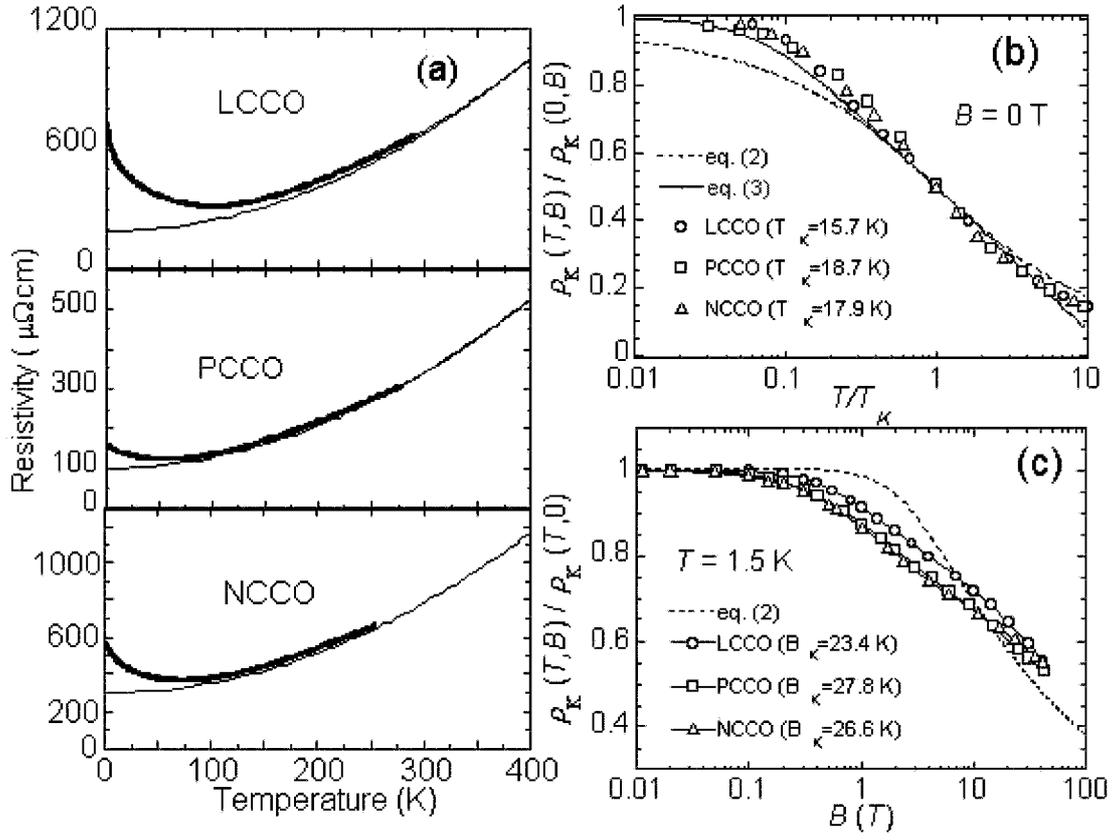

Fig. 6 / T. Sekitani et al.

FIG. 6: Fitting procedure to obtain $\rho_K$ by subtracting $\rho_0 + \rho_i$ from the experimental resistivity (a), and resultant $\rho_K(T, B = 0\text{ T})$ (b) and $\rho_K$ ($T = 1.5$ K, $B$) (c). The $T$ and $B$ dependence of $\rho_K$ are compared with the theoretical predictions from single-impurity Kondo scattering: the dashed lines represent the KMHZ or generalized Hamann formula (eq. (2)) and the solid line represents the empirical formula (eq. (3)) with $T_K = 15.7$ K, 18.7 K, and 17.9 K for LCCO, PCCO, and NCCO, respectively.